# Mapping of Electronic Band Gap along the Axis of Single InAs/InSb$_x$As$_{1-x}$ Heterostructured Nanowire

Atanu Patra, Monodeep Chakrobarty and Anushree Roy[*]

*Department of Physics, Indian Institute of Technology Kharagpur, Kharagpur 721 302, India.*

**Abstract:** We report the graded electronic band gap along the axis of individual heterostructured WZ-ZB InAs/InSb$_{0.12}$As$_{0.88}$ nanowires. Resonance Raman imaging has been exploited to map the axial variation in the second excitation gap energy ($E_1$) at the high symmetry point ($L$ point) of the Brillouin zone. We relate the origin of the observed evolution of the gap energy to the fine tuning of the alloy composition from the tip towards the interface of the nanowire. The electronic band structures of InAs, InSb and InSb$_x$As$_{1-x}$ alloy systems at $x$=0.125, 0.25, 0.50, 0.75 and 0.875, using all electron density functional theory code Wien2k, are reported. The measured band gap along the axis of the InAs/InSb$_{0.12}$As$_{0.88}$ nanowire is correlated with the calculated gap energy at the A point and the L point of the Brillouin zone for InAs and InSb$_{0.125}$As$_{0.875}$, respectively. We draw a one-to-one correspondence between the variation of the $E_1$ gap and the fundamental $E_0$ gap in the calculated electronic band structure and propose the graded fundamental gap energy across the axis of the nanowire.

---

[*] Email: anushree@phy.iitkgp.ernet.in

The advancement of nanowire (NW) growth techniques opens a new horizon in the field of electronic band engineering. The fine tuning of the electronic band gap achieved by varying the composition[1] or strain in the crystal structure[2,3,4] in a single III-V semiconductor NW is of immense interest from both physics and application view points. For example, it has been shown that light emission from GaAs of wurtzite (WZ) phase can be controlled by varying external uniaxial stress on the NW.[4] On the other hand, the possible characteristics of InSb and $InSb_xAs_{1-x}$ as topological insulators under stress or varying composition (*x*), has enhanced the interest in these systems from the physics point of view.[5,6]

The efficiency of InAs or InSb-based semiconductors to emit and detect long wavelength radiation harnesses the potential use of these systems in the fabrication of infra-red light sources (like lasers and LEDs) or detectors in optical communications and sensors. The need of finding an alternative of the most commonly used HgCdTe alloys as mid-infrared detectors, prompted researchers to investigate narrow band gap InSb and $InSb_xAs_{1-x}$ ternary alloy systems.[7] The InSbAs-based NWs are new state of art systems, which, in recent times have been of special interest due to the unavoidable modulation of alloy composition, axial strain[8,9,10] or defect states,[11] induced during the growth of a NW. All these factors are expected to fine tune the band gap along its axis. The above mentioned studies are the incentives for us to study the variation of the band gap in individual InAs/InSbAs heterostructured NW.

In this work, we have demonstrated graded band gap in InAs/InSbAs NWs. The narrow band gap of InSbAs alloy, renders it difficult to study this system using conventional luminescence measurements using visible light as the excitation energy, a most commonly used technique to probe the electronic band gap of semiconductors. We report the resonance Raman (RR) mapping on individual InAs/$InSb_{0.12}As_{0.88}$ NW, which demonstrates the variation of the $E_1$ gap (next higher gap than the fundamental $E_0$ gap) along its axis. We have correlated the origin of the observed tuning in the band gap to the fine modulation of the alloy composition along the axis of the NW. For a better understanding of the experimental findings, *ab initio* band structure calculations have been performed for various compositions (*x*) of $InSb_xAs_{1-x}$ alloy using the density functional code Wien2k. Apart from the pristine structures, we have employed supercell to accommodate the varying alloy concentrations (*x*=0.125, 0.25, 0.50, 0.75, 0.875). The composition $InSb_{0.125}As_{0.875}$, used in one of our calculations, is in close proximity to the same of InSb$_{0.12}$As$_{0.88}$, studied experimentally and hence, facilitates a good comparison between experimental and theoretical findings. Next, by establishing a one-to-one correspondence between the calculated $E_1$ and $E_0$ band gaps, we map the fundamental gap along the axis of the NW.

**RESULTS AND DISCUSSION**

**Raman image of a single nanowire:** Figure 1(a) shows the SEM image of the InAs/InSb$_{0.12}$As$_{0.88}$ heterostructured NWs. The image is recorded at an angle of 45° with (111) plane (growth direction of the NW). The length (*l*) and diameter (*d*) of the NW for InAs and InSbAs sections are *l*=0.94 µm, *d*=70 nm and *l*=0.96 µm, *d*=82 nm, respectively. Figure 1(b) presents the SEM image of a single NW, in which the InAs stem and InSbAs alloy segments are marked. This article exploits Raman mapping or imaging as a probe to study individual NWs. Raman imaging is a non-invasive technique based on Raman spectra collected at each point of a raster-scanned sample. The false color image, based on the intensity of characteristic Raman modes of materials, is then generated. Hence, this technique reveals the map of the phonon dynamics of the solid under scan. The Raman mapping of a single InAs/InSb$_{0.12}$As$_{0.88}$ NW, obtained by plotting the integral intensity near the spectral domain of the TO modes of InSb and InAs, are shown in Fig. 1(c) and (d), respectively. The intermixing of InAs-like and InSb-like signals in the alloy segment results in a color contrast between the InAs stem and InSbAs part, as displayed in the merged image 1(e).

**Mapping of band gap by resonance Raman imaging:** For a visual representation of the variation of the band gap along the axis of the NW, we carried out RR mapping. Because of the mismatch between the energy of optical phonon and optical photon, visible electromagnetic radiation does not interact with the lattice directly to create the former. In Stokes Raman scattering, the incident light creates a phonon only via the creation of an electron-hole pair. The scattering probability enhances when the value of the excitation energy coincides with the electronic band gap and the resonance occurs. Thus, the variation of the resonance energy along the axis maps the tuning of the band gap in the heterostructured NW. We have recorded Raman image of the same NW, shown in Fig. 1 (b), using different excitation energies. The intensity of the TO mode of InAs was then estimated for all image points for all excitation wavelengths. Fig. 2 is the RR map of the NW. The colour scale corresponds to the intensity near the spectral domain of the TO mode of InAs for different excitation energies (shown on y-scale). The images of the NW for different laser excitation energies are corrected for incident laser power, the $\omega^4$ law and wavelength dependent spectral response of our experimental set up.[12] The recorded data are also corrected by taking into account the optical properties of InAs and $InSb_{0.12}As_{0.88}$ at different excitation energies. For the latter we used the empirical Vegards law[13] and optical properties of InAs and InSb from Ref.[14] We find that the response of the Raman intensity to excitation energies is similar across the InAs stem segment, however, gradually broadens towards the low energies in the alloy segment. Keeping in mind that discrete wavelengths were used in resonance mapping and far-field diffraction limits the spatial resolution of Raman measurements, the broadening of the intensity response towards the lower excitation energies, as observed in Fig. 2, is only an indication of a red-shift in maxima of the resonance energy along the axis in the alloy part. It is non-trivial to comment on the exact shift in the band gap. As the visible excitation energies, used by us, resonate with the $E_1$ gap of InAs[15] or InSb,[16] we attribute the observed change in resonance energy in Fig. 2 to the modulation of the $E_1$ gap energy along the axis of the NW.

**Origin of variation in band gap along the axis:** To investigate the possible origin for the observed variation in the band gap of the alloy segment, as observed in Fig. 2, we refer to the normalized Raman spectra over the spectral range between 195 and 230 $cm^{-1}$ (Fig. 3(a)), recorded along the axis of the NW using 514 nm as the laser excitation wavelength. The characteristic spectra over the extended range (between 155 and 240 $cm^{-1}$) for the ones marked by ∗ in Fig. 3(a) are shown in Fig. 3(b). In this range, InSb-like mode, InAs-like TO and LO phonon modes appear at ~180, ~213 $cm^{-1}$ and ~233 $cm^{-1}$, respectively. The stem and alloy segments of the NW could be identified by the appearance of the InSb-like TO mode only for the latter. All spectra in the stem part, over the range between 155 and 240 $cm^{-1}$, are fitted with two Lorentzian functions for InAs-like TO and LO modes. The spectra, recorded in the alloy segment, are fitted with three Lorentzian functions, the extra one is for the InSb-like TO mode. The deconvoluted components are shown by the blue dashed lines in Fig. 3(b) and the net fitted spectra are shown by the solid lines. The spectra, for which both InAs-like and InSb-like modes appeared, the net fitted lines are shown by the red colour. Other spectra, for which only InAs-like modes are observed, the net fitted lines are shown by the green colour. In Fig. 3(a) the magenta and blue dashed lines mark the shift in InAs-like TO mode along the axis of the NW, in the stem and alloy segments, respectively.

In Fig. 4 we plot the change in Raman shift of the InAs-like TO mode along the axis of the NW. It is to be recalled that length of the WZ-InAs stem is 0.94 µm. The spatial step size (150 nm) of Raman mapping and the far field diffraction of light limit us to mark the spectrum exactly at the interface. Within this constraint of spatial inaccuracy of measurements, the shaded area separates the InAs stem (left) and InSbAs alloy segment (right). We mark the starting point of the alloy section (at 1.05 µm, confirmed by the appearance of InSb-TO mode) by the red dotted line in Fig.4. The magenta and blue dashed lines are guide to the eyes to the data points in these two parts. We find that within our experimental error bar, the Raman shift is nearly constant over the stem region, however, decreases over the alloy segment from the near-interface side towards the tip. As the change in Raman shift in the alloy segment is only by 1.2±0.3 cm$^{-1}$, to check the reproducibility of the results, the experiments have been repeated several times on different NWs.

In reference to Fig. 3 and Fig. 4 we discuss below the possible explanations for the observed modulation in the band gap along the axis of the NW.

The band gap in a one dimensional NW system is expected to increase with a decrease in its diameter due to confinement of the charge carriers.[17,18] Such effect is observed in NWs with diameter below 20 nm. The diameter of the NWs, under study, is too large to exhibit a shift in the band gap due to electron confinement. Furthermore, the tapering of the NWs, under study (see Fig. 1(b)), would have increased the band gap of the NW towards the tip. This is contrary to what we observe in Fig. 2, and hence, such a possibility can be eliminated.

The CuPt-type ordering in the crystal structure of a III-V semiconductor alloy decreases the band gap.[19] For CuPt-type ordering in III-V alloy, As-rich and Sb-rich group V layers alternate in the crystal structure. As this doubles the unit cell, one would expect either an appearance of new phonon modes or an increase in phonon spectral width due to such disorder in the crystal structure.[20] In Raman spectra, shown in Fig. 3, we do not observe any new Raman mode or broadening of the spectra, recorded along the axis of InSbAs segment. Moreover, such disorder in the crystal structure near the surface of the NW, as observed in Ref.,[8] would not result in a gradual decrease in the Raman shift towards the tip, as shown in Fig. 4. From the above discussion it is reasonable to believe that we do not observe any reflection of the CuPt ordering in Raman line profile of NWs. Thus, we rule out such disorder in the crystal structure as the origin of the observed decrease in the band gap in Fig. 2, reflected from RR mapping.

Keeping in mind that the heating effect can also result is shift in the band gap, we recorded both Stokes and anti-Stokes Raman spectra along the axis of the NWs [Fig. S1 in the supplementary section]. We observed that the intensity ratio of the InAs-like TO mode ($I_S/I_{AS}$) remains nearly unchanged along the axis, confirming that change in temperature is not the origin of the observed change in the band gap along the axis of the NW, as shown in Fig. 2.

A strain in the crystal structure can result in a change in the band gap.[2,3,4] It is well-known that a strain also causes a shift in the frequency of the phonon modes. Using DFT calculations, for the optimized crystal structure of WZ-InAs, the lattice parameters are estimated to be $a=$ 4.306 Å and $c=$7.031 Å. These calculated values are very close to the reported values of the same in the literature.[21] In addition, we also estimate the lattice constant of ZB-InSb$_{0.125}$As$_{0.875}$ as $a=$6.144 Å. Due to lattice mismatch between InAs and InSbAs alloy, one would expect a compressive strain in the alloy segment at the stem-alloy interface. A relaxation of the compressive strain along the axis of the NW is expected[22] to result in a red shift of the Raman wavenumber and hence, a decrease[4] in the band gap towards the tip. This is what we find in Fig. 2 and 4. Nonetheless, it is generally believed that the strain due to the lattice mismatch relaxes within 20 nm at the heterointerface. This rules out the origin of the observed evolution of the Raman shift in Fig. 4 to be the strain in the crystal structure. Thus, we believe that the interfacial strain and its subsequent relaxation towards the tip may not be the origin of the observed shift in the band gap in Fig. 2.

Another possible origin of the observed shift in the band gap is the variation in the composition $x$ along the axis of the NW. Let us refer to the inset of Fig. 4, which plots the expected variation of the Raman shift of InAs-like and InSb-like TO mode with composition of the alloy (Sb fraction, $x$) as obtained from Ref.,[23,24] and also uses our measured value of the Raman shift of bulk ZB-InAs ZB-InSb TO mode at 218 and 180 cm$^{-1}$. We find that for x=0.12, the InAs-like TO mode in the InSb$_x$As$_{1-x}$ alloy is expected to appear at 214.1 cm$^{-1}$. As shown in Fig. 4, above the stem-alloy interface (*i.e.*, above the shaded area at 1.05 μm) the same appears at 213.8 ±0.3 cm$^{-1}$. Within our experimental accuracy, this measured Raman shift of the InAs-like

TO mode is same as expected for the $InSb_{0.12}As_{0.88}$ alloy. Since the electronic band structure of any material is near-sighted,[25] we can safely conclude that the effect of the interface will not be seen at 1.05 µm from the bottom of the InAs stem (it is to be recalled that the length of the stem is 0.94 µm, as measured from SEM image). The further shift of the InAs-like TO phonon mode towards the tip by 1.2 cm$^{-1}$ (as observed in Fig. 4 above the shaded area) corresponds to an increase in $x$ by 0.06.

The exact origin for the variation in alloy composition along the axis of the NW is unknown to us. However, we find that for II-VI NWs, there are quite a few reports in the literature in which the temperature gradient in the axial direction during the growth process has been attributed to the observed gradual change in composition along the axis.[26,27] Moreover, during the CBE growth of NWs the Au seed under the In flux forms AuIn alloy.[28] It has been shown by Tsai and William[28] that the reaction of $AuIn_2$ with Sb is more favourable than with As. For the Au-In-As ternary, only five Au-In phases are involved. In contrary, eight Au-In phases are available for the Au-In-Sb. This may result in a slight increase in Sb fraction during the growth process. We emphasize that the small change in $x$, which could be measured by Raman mapping, is very small and below the sensitivity of EDX technique, generally used for the detection of the composition of the NW.

**Graded fundamental band gap along the axis of the NWs:** As mentioned earlier, by RR imaging we mapped the $E_1$ gap along the axis of the NW. There are quite a few reports in the literature in which the electronic band structure near the fundamental band gap ($E_0$ gap) of InAs, InSb or their alloy systems have been studied extensively. To the best of our knowledge a systematic study on the $E_1$ gap of InSbAs (which we measure experimentally by RR measurements) is still lacking in the literature. To estimate the variation of the $E_1$ gap for different $x$ in $InSb_xAs_{1-x}$ systems, we calculate the electronic band structure of the alloy at different $k$ points of the Brillouin zone for $x$=0, 0.125, 0.25, 0.50, 0.75, 0.875 and 1.0. The parent materials, (InAs and InSb), have the space group of *F43m* (216) with no inversion symmetry. The supercell structures, which result on doping the parent unit cell (*F43m*), acquire tetragonal (for $x$= 0.125, 0.50 and 0.875) and primitive cubic structures (for $x$= 0.25 and 0.75). Thus, the ternary alloys are treated as pseudo binary materials, where As is progressively replaced by Sb. The optimized lattice constants of $InSb_xAs_{1-x}$ are obtained from Birch-Murnaghan[29] fitting of energies with volume and they are in excellent match with other similar calculations.[30]

The calculated electronic structures at different $k$ points of BZs of the compounds for $x$=0, 0.125, 0.25, 0.50, 0.75, 0.875 and 1 are shown in Fig.5. The unit cell of the parent and the alloy compounds, which we considered in our calculations, are also shown at the bottom in Fig.5. As the stem of the NW, investigated experimentally, has the WZ crystal structure, Fig. 5 also includes the band structure and the unit cell of WZ-InAs. The $E_1$ gap in each band structure is shown by the red arrow. The estimated variation of the $E_1$ and the $E_0$ band gap with $x$, thus obtained, are shown in Fig. 6(a) and (b). We exploit measured value of change in $x$ by 0.06 in the alloy segment towards the tip, as obtained from the shift in Raman spectra in Fig. 4, to estimate the expected variation in band gap along the axis of the NW, as could be proposed from DFT calculations. In the inset of Fig. 6 we plot the magnified view in the range of $x$ between 0 and 0.25 over which the $E_1$ gap (upper panel) varies between 2.38 and 1.71 eV and the $E_0$ gap (lower panel) varies between 0.48 and 0.22 eV. The gradual increase in $x$ by 0.06 (from 0.12 to 0.18) in the alloy part towards the tip is expected to map the $E_1$ gap from 2.08 eV to 1.90 eV (shown by the green-shaded area in Fig. 6(a)). We draw a one-to-one correspondence between calculated $E_1$ gap and $E_0$ gap (orange-shaded area) over the given range of $x$ in Fig.6. With this information from our DFT calculations and measured variation of $x$ in the alloy segment we propose the expected grading of the fundamental band gap over the range between 0.34 eV and 0.29 eV along the axis of the NW in the alloy segment and show it in the inset of Fig. 6(b). The $E_1$ and $E_0$ band gaps of WZ-InAs in the stem are calculated to be 2.31 and 0.58 eV.

**CONCLUSION**

In this article, we report the graded band gap along the axis of InAs/InSb$_{0.12}$As$_{0.88}$ NW. RR imaging maps the modulation of the $E_1$ gap of the system. We have correlated the observed band gap energy to the change in composition of the alloy from the interface towards the tip. We have calculated electronic band structures of the parent and alloy compounds using FP-LAPW based DFT package Wien2k. By correlating $E_0$ and $E_1$ gaps as a function of composition from our calculated band structure, we can gain an insight about the fundamental gap ($E_0$) variation along the axis of the NW.

Furthermore, in this article we report the extended band structure for the whole family of the InSb$_x$As$_{1-x}$ alloy. Thus, the resonance Raman imaging technique along with the calculated band profile can be used as an easy characterizing tool to probe the smooth grading of the band gap in this system. We believe that the fascinating graded band gap in III-V NWs will find its potential applications in fabricating optoelectronic devices *e.g.,* nanoscale multi-terminal photo detectors.

**METHOD**

**Growth of NWs:** Axial InAs/InSb$_x$As$_{1-x}$ heterostructured NWs were grown on InAs(111)B substrate by chemical beam epitaxy (CBE) in a Riber Compact-21 System. The gold (Au) film was deposited on InAs(111)B wafers by thermal evaporation in a separate evaporator chamber and then was transferred to the CBE system. The tertiarybutylarsine (TBAs) flow was continued for 20 min and the CBE chamber temperature was maintained at 500±5 °C to remove the surface oxide and to dewet the Au film into nanoparticles. The InAs stem was grown at a temperature of 420±5 °C when the metallorganic (MO) line pressure was kept fixed at 0.3 and 1 Torr for 60 min and 75 min for trimethylindium (TMIn) and (TBAs) respectively. To grow the InSbAs alloy segment, the tris(dimethylamino)antimony (TDMASb) precursor with the pressure 0.37 Torr was switched on and TBAs precursor line was maintained at 0.50 Torr. During this process, un-interrupted TMIn flux as well as the chamber temperature was maintained. TDMASb precursor was kept on for 60 min. To the end, the TMIn flux was stopped and the sample was cooled down under TDMASb and TBAs flow.

**Characterization of the NWs:** For detail structural characteristics of these NWs see Ref..[11] The composition of the NWs was estimated using energy dispersive x-ray spectroscopy (EDX) in a JEOL 2200FS microscope, working at 200 keV. For NWs, under study, the value of *x* is reported to be 0.12. The InAs stem and InSbAs alloy segments were found to be in wurtzite (WZ) and zinc-blende (ZB) phase, respectively. FE-SEM image of single NW was measured using MERLIN (make Zeiss, Germany) scanning electron microscope at 5 keV.

**Raman mapping of the NWs:** The Raman mapping, by following the point scanning technique, was carried out in the backscattering configuration on individual InAs/InSb$_{0.12}$As$_{0.88}$ NW, transferred mechanically onto a silicon substrate. A computer controlled motorized XY stage was used for scanning. The scanning-step size was 150 nm. The Ar$^+$-Kr$^+$ laser (Model

Innova 70C Spectrum, Make Coherent, USA) with eight different excitation energies between 1.919 and 2.607 eV was used for RR measurements. The laser light was focused on the NW with a 100×objective lens (numerical aperture 0.90) and the excitation power was kept at 50 $\mu$W. We have used the Raman spectrometer T64000 (JY, France) in the subtractive mode with a step spectral resolution of 0.3 cm$^{-1}$. The investigation was carried over several individual NWs and reproducible results could be obtained.

**DFT calculations:** All theoretical calculations were performed with full potential linearized augmented plane wave (FP-LAPW) method,[31,32] implemented within WIEN2k package.[33] The basis set convergence parameter ($R^{min}_{MT}K_{max}$) was set to 9. The spherical harmonic functions inside the muffin-tin spheres was limited by $l_{max}$=12, where muffin tin radii for In, As and Sb were kept fixed at 1.9, 2.1 and 2.15, respectively. In interstitial regions the charge density and potential were defined through $G_{max}$ at 14 Bohr$^{-1}$. The tetrahedron method was employed for BZ integrations within the self-consistency cycles. The band structures were computed on a discrete $k$ mesh along high-symmetry directions. We applied GGA functional to calculate the structural properties. Next, we used the mBJLDA potential along with spin-orbit coupling for band structure calculations. The force on each atom was adjusted to relax less than 0.5 mRyd/a.u.

**Acknowledgements.** AT and AR thank Central Research Facility of IIT Kharagpur, for the FE-SEM image of the NWs and also for the availability of the Raman spectrometer. Auhors acknowledge the use of the computing facility from the DST-Fund for Improvement of S&T infrastructure (phase-II) Project installed in the Department of Physics, IIT Kharagpur, India. Authors also thank Professor Lucia Sorba and her group at NEST-Istituto Nanoscienze-CNR,Pisa in Italy for the NW growth and valuable discussion.

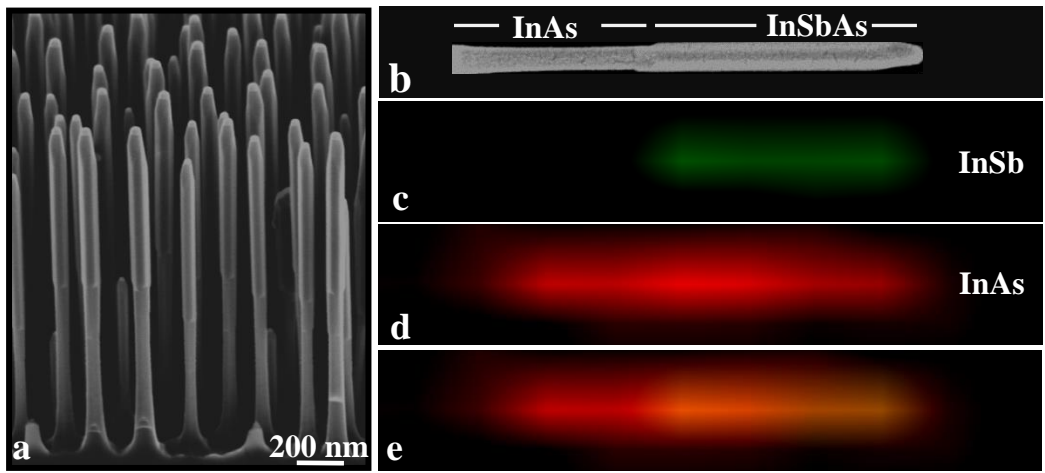

FIG. 1: (a) 45° tilted SEM image of aligned InAs/InSb$_{0.12}$As$_{0.88}$ NWs. (b) SEM image of a single NW. Characteristic Raman image of an individual NW with 514 nm as excitation wavelength showing (c) InSb-like and (d) InAs-like signal along the axis of the NW. For clarity, the intensity of InSb-like signal is magnified 3 times. (e) Merged image showing intermixing of (c) and (d) for the alloy segment of the NW.

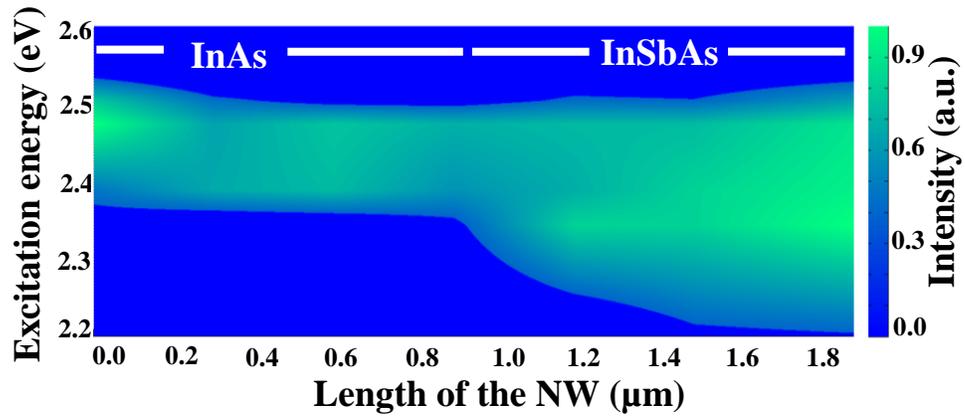

**FIG. 2:** The contour plot demonstrating the intensity of InAs-like TO mode for different excitation energy along the axis of the NW. The colour scale represents the intensity of the Raman mode.

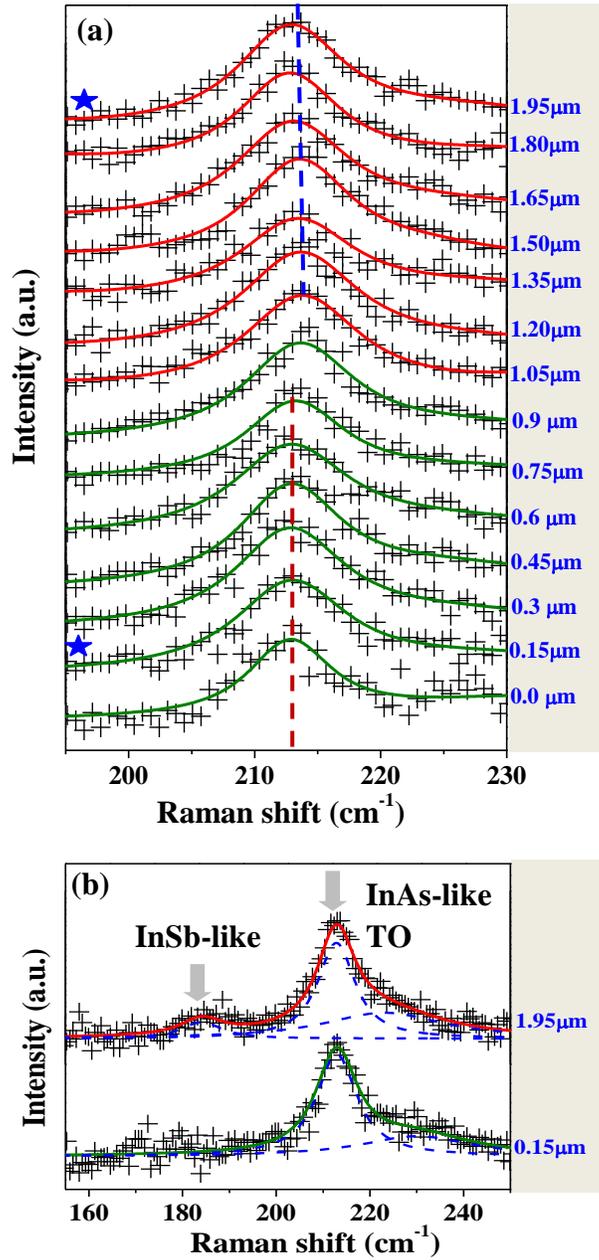

FIG. 3: (a) Raman spectra along the axis of the NW with 514 nm as the excitation wavelength. (+) symbols are experimental data points. The blue and red dashed lines mark the shift in InAs-like TO mode in alloy and stem segments. The scanning steps are marked on right. (b) Spectra over the extended range for the ones marked by blue star in (a). For the measured spectra, in which only InAs-like modes were observed, the net fitted spectra are shown by green lines. For the measured spectra, in which both InSb like TO mode InAs-like modes were observed, the net fitted spectra are shown by red lines.

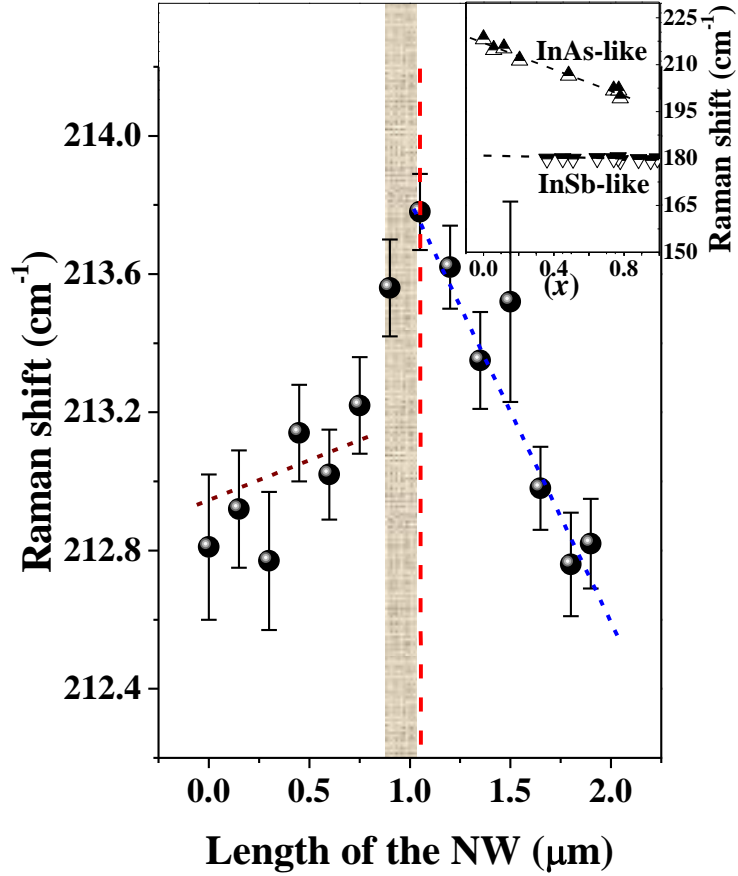

Fig. 4 The change in Raman shift of InAs-like TO mode along the axis of the NW. The red dashed line separates the InAs stem to the left and InSbAs alloy segment to the right. The magenta and blue dashed lines are guide to the eyes to the data points in these two parts of the NWs. Inset of the figure shows the variation in Raman shift of InAs-like and InSb-like TO phonon mode with Sb fraction ($x$), as obtained from Ref. [23,24] and our measured values of the Raman shift of the TO phonon mode of pristine ZB-InAs.

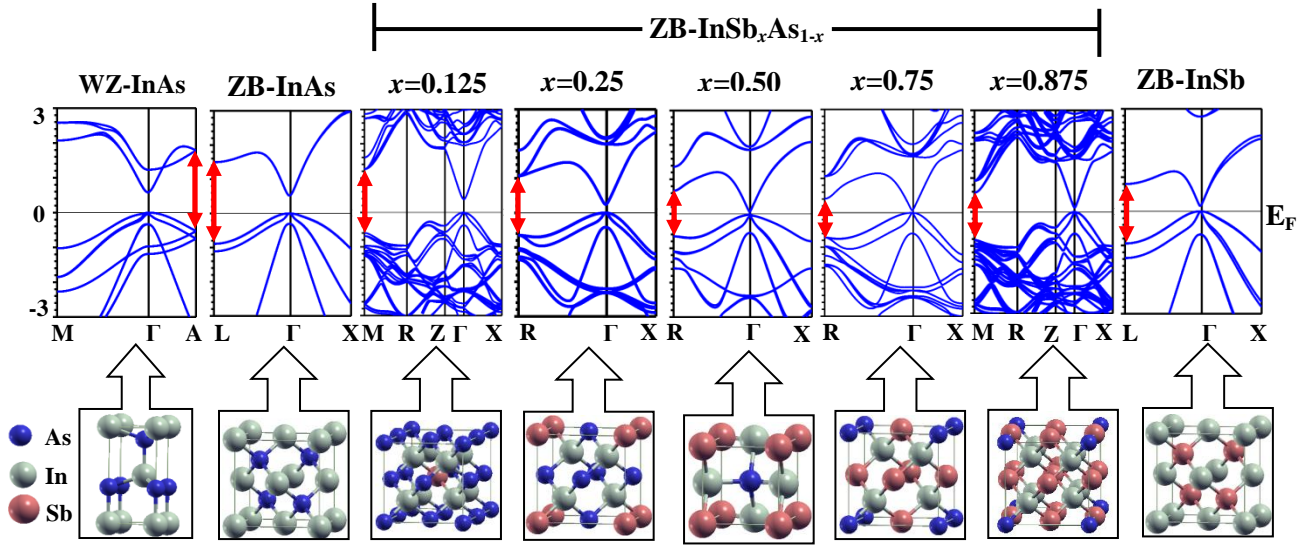

**FIG. 5 :** Calculated electronic band structure of ZB-InAs, ZB-InSb$_x$As$_{1-x}$ and ZB-InSb. The $E_1$ gaps are marked by red arrow. The corresponding unit cells are shown below.

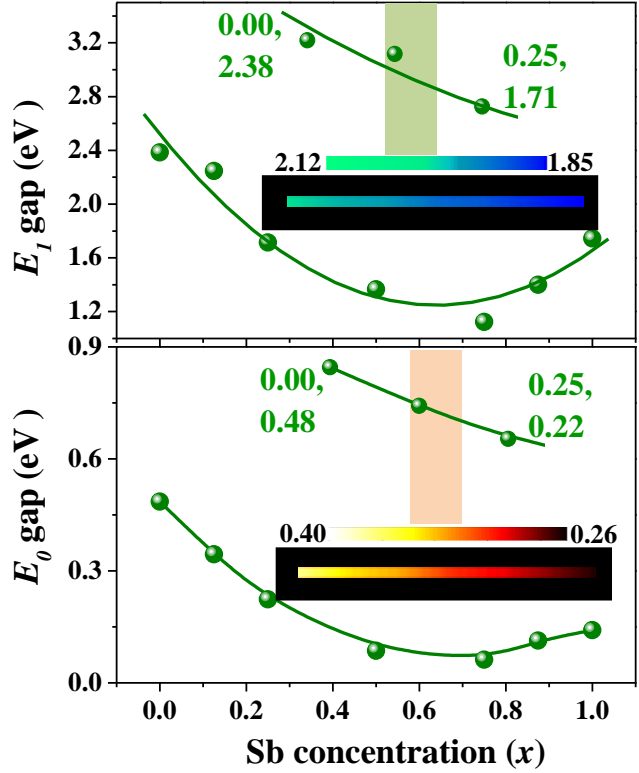

FIG.6: (a) Calculated variation of (a) $E_1$ gap and (b) $E_0$ gap with $x$ in ZB-InSb$_x$As$_{1-x}$. The plots over the range from $x=0.00$ to $x=0.25$ are shown in the inset. The variation in the $E_1$ gap (in eV), due to measured change in $x$ from 0.12 to 0.18, is shown by the green shaded area in the inset of (a). The corresponding expected variation in the fundamental gap is shown in the panel (b) by the red shaded area. The expected graded $E_1$ and $E_0$ band gaps (in eV) along the axis of the alloy segment in the NW are also shown in the corresponding panels.

# REFERENCES

1. Yan, R.; Gargas, D.; Yang, P. Nanowire Photonics. *Nat.* **2009**, 3, 569-576.

2. Skold, N.; Karlsson, L. S.; Larsson, M. W.; Pisto, M. E.; Seifert, W.; Tra1gardh, J.; Samuelson, L. Growth and Optical Properties of Strained GaAs-Ga$_x$In$_{1-x}$P Core-Shell Nanowires. *Nano. Lett.* **2005**, 5, 1943-1947.

3. Signorello, G.; Karg, S.; Bjork, M. T.; Gotsmann, B.; Riel, H. Tuning the Light Emission from GaAs Nanowires over 290 Mev with Uniaxial Strain. *Nano. Lett.* **2013,** 13, 917-924.

4. Signorello, G.; Lörtscher, E.; Khomyakov, P. A.; Karg, S.; Dheeraj, D. L.; Gotsmann, B.; Weman, H.; Riel, H. Inducing a Direct-to-Pseudo Direct Bandgap Transition in Wurtzite GaAs Nanowires with Uniaxial Stress. *Nat. Commun*. **2014,** 5, 3655.

5. Feng, W.; Zhu, W.; Weitering, H. H.; Stocks, G. M.; Yao, Y.; Xiao, D. Strain Tuning of Topological Band Order in Cubic Semiconductors. *Phys. Rev. B* **2012**, 85, 195114.

6. Namjoo, S.; Rozatian, A. S. H.; Jabbari, I. Influence of Lattice Expansion on the Topological Band Order of InAs$_x$Sb$_{1-x}$ (X = 0, 0.25, 0.5, 0.75, 1) Alloy*s*. *J. Alloys Compd*. **2015**, 628, 458-463.

7. Svensson, J.; Anttu, N.; Vainorius, N.; Borg, B. M.; Wernersson, L.E. Diameter-Dependent Photocurrent in InAsSb Nanowire Infrared Photodetectors. *Nano. Lett*. **2013**, 13, 1380-1385.

8. Ercolani, D.; Rossi, F.; Li, A.; Roddaro, S.; Grillo, V.; Salviati, G.; Beltram, F.; Sorba, L. InAs/InSb Nanowire Heterostructures Grown by Chemical Beam Epitaxy. *Nanotech*. **2009**, 20, 505605.

9. Mata, M. d.; Magen, C.; Caroff, P.; Arbiol, J. Atomic Scale Strain Relaxation in Axial Semiconductor III-V Nanowire Heterostructures. *Nano.Lett*. **2014,** 14, 6614-6620.

10. Patra, A.; Panda, J. K.; Roy, A.; Gemmi, M.; David, J.; Ercolani, D.; Sorba, L. Mapping of Axial Strain in InAs/InSb Heterostructured Nanowires. *Appl. Phys. Lett*. **2014,** 14, 093103.

11. Ercolani, D.; Gemmi, M.; Nasi, L.; Rossi, F.; Pea, M.; Li, A.; Salviati, G.; Beltram, F.; Sorba, L. Growth of InAs/InAsSb Heterostructured Nanowires. *Nanotech.* **2012**, 23, 115606.

12. Loudon, R. Theory of the Resonance Raman Effect in Crystals. *J. Phys*, **1965,** 26, 677-683.

13. Vegard, L. Die Konstitution der Mischkristalle und die Raumfüllung der Atome. Z. Phys. **1921**, 5, 17–26.

14. Adachi, S. Optical Dispersion Relations for GaP, GaAs, GaSb, InP, InAs, InSb, Al$_x$Ga$_{1-x}$As, and In$_{1-x}$Ga$_x$ As$_y$P$_{1-y}$. *J. Appl. Phys*. **1989,** 66, 6030-6040.

15. Caries, R.; Saint-Cricq, N.; Renucci, J.B.; Zwick, A.; Renucci, M. A. Resonance Raman Scattering in InAs Near the E$_1$ Edge. *Phys. Rev. B* **1980**, 22, 6120-6126.

16. Menendez, J.; Vina, L.; Cardona, M.; Anastassakis, E. Resonance Raman scattering in InSb: Deformation Potentials and Interference Effects at the $E_1$ Gap. *Phys. Rev. B* **1985**, 32, 3966-3973.

17. Allan, G.; Niquet, Y. M.; Delerue, C. Quantum Confinement Energies in Zinc-Blende III-V and Group IV Semiconductors. *Appl. Phys. Lett.* **2000**, **77**, 639-641.

18. Yu, H.; Li, J.; Loomis, R. A.; Wang, L. W.; Buhro, W. E. Two- Versus Three-Dimensional Quantum Confinement in Indium Phosphide Wires and Dots. *Nat. Mater.* **2003**, 2, 517- 520.

19. Wei, S. H.; Zunger, A. Fingerprints of Cupt Ordering in III-V Semiconductor Alloys: Valence-Band Splittings, Band-Gap Reduction, and X-Ray Structure Factors. *Phys. Rev. B* **1998**, **57**, 8983-8988.

20. Cheong, H. M.; Mascarenhas, A, Effects of Spontaneous Ordering on Raman Spectra of GaInP$_2$. *Phys. Rev. B* **1997**, 56, 1882-1887.

21. Kriegner, D.; Panse, C.; Mandl, B.; Dick, K. A.; Keplinger, M.; Persson, J. M.; Caroff, P.; Ercolani, D.; Sorba, L. ; Bechstedt, F.; Stangl, J.; Bauer, G. Unit Cell Structure of Crystal Polytypes in InAs and InSb Nanowires. *Nano Lett.* **2011**, 11, 1483-1489.

22. Cerdeira, F. Buchenauer, C. J.; Pollak, F. H.; Cardona, M. Stress-Induced Shifts of First-Order Raman Frequencies of Diamond- and Zinc-Blende-Type Semiconductors. *Phys. Rev. B* **1972**, 5, 580-593.

23. Cherng, Y. T.; Ma, K. Y.; Stringfellow, G. B. Raman Scattering in InAs$_{1-X}$Sb$_x$ Grown by Organometallic Vapor Phase epitaxy. *Appl. Phys. Lett.* **1988**, 53, 886-887.

24. Huang, L.; Li, Z. F.; Chen, P. P.; Zhang, Y. H.; Lu, W. Far Infrared Reflection Spectra of InAs$_x$Sb$_{1-X}$ (X=0-0.4) Thin Films. *J. Appl. Phys.* **2013,** 113, 213112.

25. Prodan, E; Kohn, W. Nearsigthedness of Electronic Matter. *PNAS*. **2005,** 102, 11635-11638.

26. Zhuang, X.; Ning, C. Z.; Pan, A. Composition and Bandgap-Graded Semiconductor Alloy nanowires**.** *Adv. Mater.* **2012**, 24, 13-33.

27. Pan, A.; Zhou, W.; Leong, E. S. P.; Liu, R.; Chin, A. H.; Zou, B.; Ning, C. Z. Continuous Alloy-Composition Spatial Grading and Superbroad Wavelength-Tunable Nanowire Lasers on a Single Chip. *Nano Lett.* **2009**, 9, 784-788.

28. Tsai, C. T.; Williams, R. S. Solid phase Equilibria in the Au-Ga-As, Au-Ga-Sb, Au-In-As,and Au-In-Sb Ternaries. J. Mater. Res. **1986**,1, 352-360.

29. Murnaghan, F. D. The Compressibility of Media Under Extreme Pressures, *Proc. Natl. Acad. Sci*.**1944**, 30,244-247.

30. Namjoo, S.; Rozatian, A. S.H.; Jabbari, I.; Puschnig, P. Optical Study of Narrow Band Gap InAs$_x$Sb$_{1-x}$ (x = 0, 0.25, 0.5, 0.75, 1) Alloys. *Phys. Rev. B* **2015**, 91, 205205.

31. Singh, D.; Nordstrom, L.; Plane Waves, Pseudo Potentials, and the LAPW Method, Springer, Berlin (1994).

32. Madsen, G. K. H.; Blaha, P.; Schwarz, K.; Sjostedt, E.; Nordstrom, L. Efficient Linearization of the Augmented Plane-Wave Method. *Phys. Rev. B* **2001**,64, 195134.

33. Blaha, P.; Schwarz, K.; Madsen, G.; Kvasnicka, D.; Luitz, J. WIEN2 K, An Augmented Plane Wave + Local Orbital Program for Calculating Crystal Properties; Karlheinz Schwarz, Technical University: Vienna, Austria, 2001.